# From hollow gold nanoparticles to gold nanorings: Morphological tunability of the plasmonic response

*Martin Prieto[a,$], Raul Arenal[*b,c,$], Luc Henrard[d,$], Leyre Gomez[a,e], Victor Sebastian[a,e], Manuel Arruebo[*,a,e]*

[a]Department of Chemical Engineering and Aragon Nanoscience Institute (INA), EdificioI+D+i, C/Mariano Esquillor s/n, University of Zaragoza, 50018 Zaragoza, Spain.

[b]Laboratorio de Microscopias Avanzadas, Instituto de Nanociencia de Aragón, Universidad de Zaragoza, 50018 Zaragoza, Spain.

[c]ARAID Foundation, 50018 Zaragoza, Spain.

[d]Department of Physics, University of Namur. 61, rue de Bruxelles, 5000 Namur, Belgium.

[e]Networking Research Center on Bioengineering, Biomaterials and Nanomedicine, CIBER-BBN, 50018 Zaragoza, Spain.

*Email: arenal@unizar.es ; arruebom@unizar.es

[$] Equal contribution.

**Abstract:**

The optical and morphological properties of hollow gold nanoparticles (HGNPs) can be finely modified by employing Poly-L-Lysine hydrobromide (PLL), an homo poly aminoacid of the L-lysine, used as reducer and stabilizer. We investigate locally the plasmonic response of these nanostructures by electron energy loss spectroscopy (EELS). The plasmon excitations are interpreted by discrete dipolar approximation (DDA) simulations. We demonstrate that this controlled top-down morphological modification a fine tuning of the optical response is possible. Unlike the traditional lithographic techniques, this has been achieved in a controlled manner using wet chemistry, enabling the potential use of these nanostructures for a broad range of plasmonic applications, including biomedicine, catalysis and quantum communications.

**Introduction.**

Hollow gold nanoparticles are interesting plasmonic nanostructures showing collective electron charge oscillations named localized surface plasmons resonances (SPR) which can exhibit enhanced near-field amplitude at the resonance frequency[1]. Those nanomaterials are widely used in catalysis, medicine, sensing and photonics. In catalysis, compared to solid nanoparticles, they show enhanced catalytic properties due to the cage effect by confining the reactants inside their hollow interiors and those inner crystalline facets show a superior catalytic activity[2]. In medicine, therapeutical and diagnostic applications benefit from their plasmonic response. In therapy, they are used in cancer treatment as transductors of the absorbed near infrared light into heat causing selectively cellular death in the nanoshell-laden cells[3] or they can also be used to trigger the release of therapeutic molecules in drug[4] or gene[5] delivery applications. Those hollow structures show improved properties such as prolonged blood circulation half-life compared to other NIR absorbing nanoparticles including core/shell $SiO_2$/Au nanoparticles[6] and they also show less cytotoxicity than gold nanorods[7]. In diagnosis, they are used as photoacoustic imaging tags[8], as contrast agents in optical coherence tomography [9, 10] and in multi-photon microscopy[11]. They also can be used as biosensors[12] and thermal tags[13].

Galvanic replacement is the most common synthesis protocol used for the synthesis of those hollow gold nanoparticles, in which a sacrificial metal (i.e., Ag or Co) is oxidized at the same time that a gold (III) salt is reduced on it generating a hollow Au shell because the reduction potential of $AuCl_4^-$/Au (0.99 V vs. SHE) is more positive than that of AgCl/Ag (0.22 V vs. SHE) or $Co^{2+}$/Co (-0.28 V vs. SHE). Different reductors and/or stabilizers are used during the synthesis including polyvinylpyrrolidone (PVP)[2, 9, 14], citrate groups[3, 5], hexadecyltrimethyl ammonium bromide[15], and so on. After synthesis and for those applications in which the resulting nanoparticles could be subjected to biofouling (i.e., opsonization) pegylation is commonly carried out by means of thiol- or amino-PEG taking advantage of the strong chemical bond between gold and sulphur[16] or nitrogen[17], respectively. Guo et al.[18] reported that a single dose of those hollow nanostructures show long term stability under physiological conditions remaining within Kupffer cells 3 months after i.v. administration. However, Goodman et al.[19] showed recently that the same nanostructures were unstable and prone to fragmentation under physiological conditions when fetal bovine serum was added to the media and that acidic media destroyed the nanoparticles even after pegylation. They attributed such destruction to the presence of Ag (used in the galvanic replacement) on the surface of the hollow gold nanoparticles and to the protein corona formed around the nanoparticles which could have a destructive effect. Considering those observations we evaluated in this work the morphological changes of those nanostructures under different pHs and under the presence or absence of an amino-based reducing agent (Poly-L-Lysine Hydrobromide) and a surface protective PEG when using cobalt as sacrificial template with the goal of understanding their plasmonic response and stability under those different conditions. In addition, we investigate the way for tuning the plasmonic response of those nanoparticles as function of the pH of the reactions taking place and the presence of amines as reductive agents. All the discussions related to these results are presented elsewhere[40].

Electron energy loss spectroscopy (EELS) developed in a transmission electron microscope (TEM) is a very powerful technique for investigating the optical response of materials at local scale[20-27]. In the present work, we have studied two different kinds of Au nanostructures (gold hollow spheres and nanorings) via low-loss EELS experiments using a probe-corrected and monochromated TEM. Particular attention has been paid to the different parameters (size/aspect-ratio, shape and dielectric environment) influencing the SPR response of these materials. Our interpretation is based on discrete dipole approximation (DDA) simulations. Furthermore, the precise composition, geometry and atomic structure of the nanoparticles have been carefully studied by energy-dispersive X-ray spectroscopy (EDS) and high-angle annular dark field using high/resolution scanning TEM (HAADF-HRSTEM) measurements. Our studies demonstrate that the fine control of the optical response of these objects is possible and it can be precisely monitored by local probe measurements. While many of the nanorings' synthesis protocols are based on top-down approaches (i.e., colloidal lithography[28]); here we describe how by following a bottom-up approach it is possible to control the geometry of the transitions between hollow gold nanoparticles and nanorings as requested by other authors[29].

**Results and discussion**

We have observed a strong morphological modification of the morphology of the gold NPs as a function of the pH of an added HCl solution, see Figure 1. At pH 7, the Au nanoparticles have a stable and relatively uniform hollow spherical morphology (HRTEM of Fig. 1 (a) and (b)), but at pH 2 the shape of the NPs changes significantly, becoming a sort of distorted nano-torus/-ring or polycrystalline compact NP, see Fig. 1 (c). An intermediated step, at pH 5, shows Au sphere-like NP, formed as a sort of decomposition of the original HGNPs, Fig. 1 (b). It is already known that acidic media promotes the de-alloying of the shell when using silver as sacrificial templates which is responsible of fragmentation[19]. In addition, adding $HAuCl_4$ to the HGNPs produces the replacement of the remaining silver resulting in reconstruction and formation of pinholes and ultimately leading to fragmentation[15].

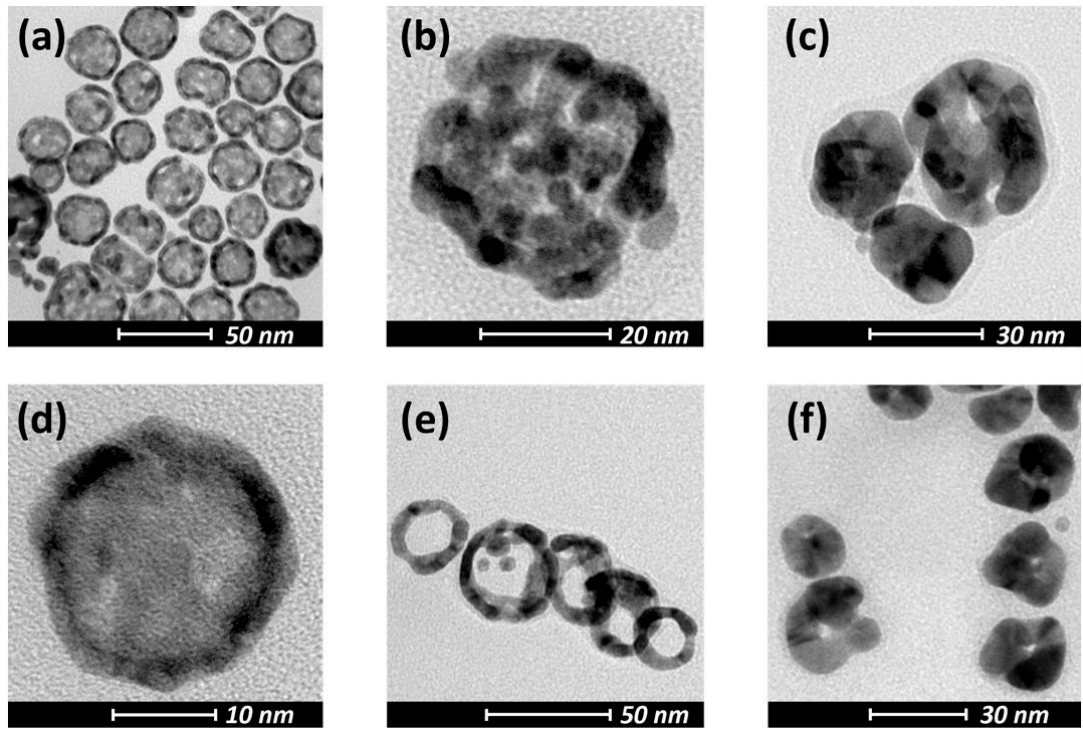

**Figure 1.** (a) – (c) Conventional TEM images characteristic of samples of aqueous dispersions of HGNPs at different pHs: 7, 5 and 2, respectively. pH modification was achieved by adding increased amounts of HCl. (d) – (f) Conventional TEM images characteristic of samples of aqueous dispersions of 1:1 Poly-L-Lysine Hydrobromide: HGNPs (0,5 mg/mL), corresponding to (d) initial, (e) after 2 h of reaction time (pH = 5) and (f) after 28 h of reaction time (pH = 5).

Upon addition of Poly-L-Lysine Hydrobromide on these HGNP, there is another very interesting morphological modification. During this treatment, the pH of the dispersion rapidly decreases to 5 due to the acidic character of the PLL with a consequent reshaping from HGNPs to gold nanorings, Fig.1 (e)-(f). As we will see below, this characteristic reshaping was analyzed over time in order to control the plasmonic response of the original HGNPs by the combination effect of the acidic environment provided by the Hydrobromide together with the reductive action of the primary amines in the side chains of the Poly-L-Lysine. Indeed, after 28 h of reaction, the nanorings (NR) underwent a morphological change and reshaped into spherical or pinhole nanoparticles with a progressive decrease in the aspect ratio of the walls of the intermediated nanoring formed after 2h of reaction time.

HAADF-HRSTEM micrographs, recorded on an Au nanoring and displayed in Fig. 2 (a) and (b), confirm the good crystalline quality of these NR.

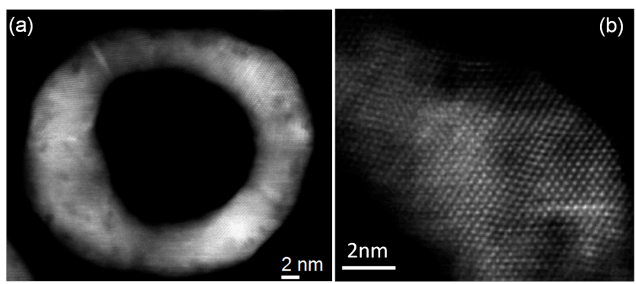

*Figure 2. (a) and (b) Two HAADF-HRSTEM micrographs recorded on an Au nanoring.*

The mechanism behind these transformations is discussed elsewhere[40].

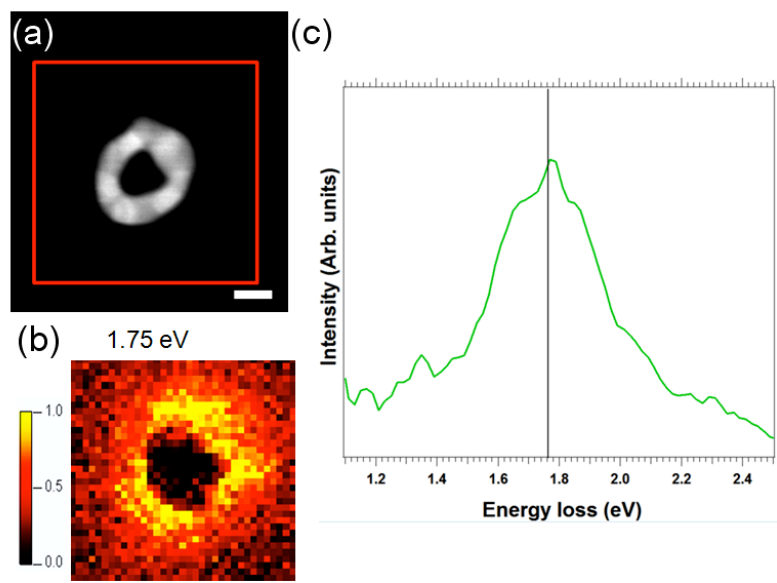

*Figure 3. (a) HAADF-STEM image (scale bar 10 nm) of an Au nanoring, where an EELS SPIMs has been recorded in the red marked area. (b) Intensity maps showing the spatial distribution of the SPR mode (1.75 eV.) for this Au NR. (c) EEL spectrum, after background subtraction, corresponding to the sum of 12 spectra. .*

As mentioned above, the local surface plasmon resonances (LSPR) depend on the shape, size and dielectric environment of the probed nano-object. We have investigated these effects and we have carried out systematic local EELS-TEM analyses on different kind of nanoparticles: hollow spheres and nanorings with different aspect-ratios/sizes. Figure 3 (a) displays the HAADF image of a NR. An EELS-SPIM has been recorded on the marked red area. A typical EEL spectrum (sum of 12 spectra) is plotted in Fig. 3 (b). This spectrum has been extracted from the green area marked on the intensity map displayed in Fig. 3 (c). The map corresponds to the intensity of the SPR mode at 1.75 eV extracted for the SPIMs after zero-loss peak (ZLP) subtraction (see Methods Section). In NR, there is a significant coupling between the two electromagnetic modes of the inner and outer surfaces of the nanostructures, leading to dipolar symmetric and antisymmetric plasmon modes.[28,34] Only the symmetric mode is observed because of its dominant excitation cross section of excitation by light or by electrons. The coupling (or hybridization) is directly related to the aspect ratio ($\sigma =$ d/0.5D, where "d" and "D" are the wall thickness and the diameter of the NR, respectively) dependence of the LSPR, as predicted and observed by optical measurements.[28] A further coupling of the 15 nm height NR with the substrate ($Si_3N_4$ film, thickness 15nm) also lead to a redshift of the plasmon modes compared to isolated self-supported NR.

We have analyzed the aspect ratio and the morphology effects on the plasmonic behaviors of the NR and HGNPs of different dimensions. A selection of EEL spectra, after ZLP subtraction and each of them corresponding to the sum of 8-12 spectra, is displayed in Fig. 4 (a). These EEL spectra have been recorded on the NR and HGNPs showed in Fig. 4 (b) and noted as (from bottom to top, (i) to (iv)) Ring 1, Ring 3, Ring 6 and Sphere 1. The geometric parameters are given in the table (Fig. 5), DDA EELS simulations[26,38] for the corresponding (same sizes/aspect ratios) NPs are presented in Fig. 4(c) and in the right column of the table. In the simulations, a $Si_3N_4$ substrate is considered. Additional experiments and simulations are provided elsewhere[40]. A very good agreement of the plasmon modes energies is found. The larger width of the experimental data is related to the unregular shape of the NR (see HAADF images).

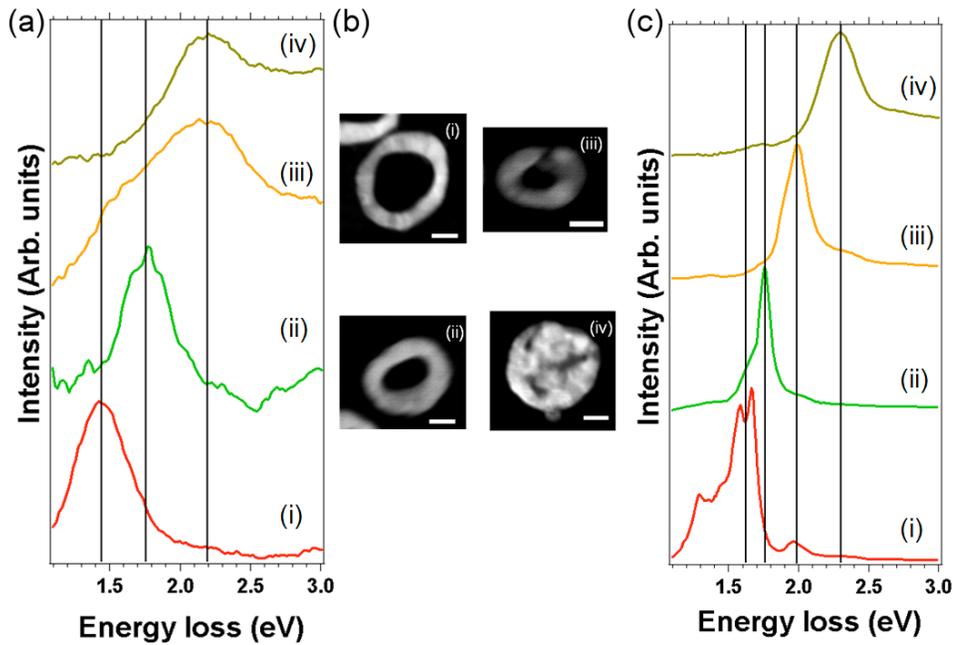

*Figure 4.* *(a) EEL spectra (each of them corresponds to the sum of 8-12 spectra, after background subtraction) recorded on nanorings of different aspect ratios (σ), (see table) ring 1 (i), ring 3 (ii), ring 6 (iii) and hollow sphere (iv). (b) HAADF images of the corresponding nanostructures. Scale bar is 10 nm for all these micrographs. (c) EELS DDA simulations for the corresponding (same sizes) NPs.*

The σ dependence is further evidenced on Fig. 5 and Fig. 6 where the experimental and simulated EELS plasmon modes energies are displayed. The σ dependence reproduces the results of Aizpurua et al.[28], i.e.: confirming that the SPR coupling is more pronounced for small σ. We note that we directly simulated EELS spectra in the present work, where only optical absorption was considered in the work of Aizpurua et al.[28] and, more importantly, tabulated 'real' dielectric data of gold has been considered rather than a Drude model. It is worth mentioning that the height of these NP (~ 15 nm) is very homogeneous for all these nanorings, and then it is not playing any effect on the observed SPR shifts.

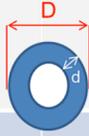

*Figure 5. Experimental data and simulations of SPR energies and morphology of several gold hollow spherical nanoparticles and Nanorings. (*) This nanoring is partially supported on the substrate (see text).*

Ring 7 ($\sigma$ = 0.8) presents a LSP resonance at 2.40 eV, which much larger than the 2 eV predicted by simulations (Fig. 5 and Fig 6). This discrepancy is explained by the fact that this nanoring is only partially supported on the $Si_3N_4$ substrate, being most part of the particle on the vacuum. Isolated NR with the same morphology with no substrate interaction present a dipolar SPR calculated at 2.35 eV, very close to the observed one.

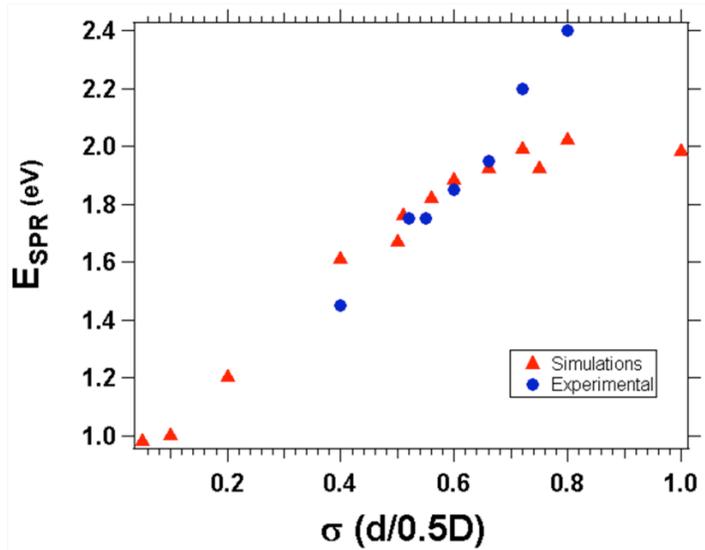

*Figure 6. Dispersion of SPR modes as a function of the aspect ratio of the Au NR and Au HGNP. Experimental data (blue circles) are those of table (Fig. 5). Simulation data (red triangles) are those for the same morphology/dimensions.*

**Conclusions**

In summary, morphological reshaping of hollow gold nanoparticles to gold nanorings or to almost spherical nanoparticles has been obtained by using PLL, which reduces the pH resulting colloidal chloroaurate ions. In addition, PLL has also a strong reductive action on the dissolved ions to produce gold nanorings or almost closed nanostructures depending on the contact time. We have monitored, by local EELS measurements, these morphological modifications and their dependence on the surface plasmon modes. In particular, we have followed the shift of the dipolar mode of the gold nanorings as a function of their aspect ratio. All these findings have been supported by DDA calculations which also helped to elucidate these observed behaviours. These results demonstrate that a fine tune of the plasmonic response of Au NP is achievable, offering the possibility of applying these NP on different fields as sensing, photonics and catalysis.

**Materials and Methods**

Materials:

Hollow Gold Nanoparticles (HGNPs) have been synthesized via galvanic replacement of cobalt with gold as previously reported by Preciado-Flores et al.[14]. All the chemicals were purchased from Sigma Aldrich and used as supplied. The solution was deareated by bubbling with argon for 40 min. 500 $\mu$L of a 1 wt.% solution of poly(vinylpyrrolidone) (PVP) with an average molecular weight of 55000 Da and 100 $\mu$L of 0.1 M sodium borohydride ($NaBH_4$) were injected under magnetic stirring to the previously deareated solution. Subsequently, 30 mL of this cobalt nanoparticle dispersion was transferred to a stirred solution of 10 mL of distilled water with 15 $\mu$L of 0.1 M $HAuCl_4$. In the case, of TEM analyses, the dispersions containing the Au NPs have been deposited on two different TEM grids: on copper carbon holey grids for some EDS and HRSTEM measurements and on $Si_3N_4$ membranes for the low-losses EELS analyses.

In some cases, the surface of the nanoparticles was functionalized by using SH-PEG (Methyl-polyethyleneglycol-thiol, 5KDa, Sigma-Aldrich)[35]. Poly-L-Lysine Hydrobromide (Mw: 15-30KDa,

Sigma-Aldrich) was stirred during 1 h with colloidal suspensions of the HGNPs (0,5 mg/mL) in 1:1, 1:2 and 1:4 proportions to allow electrostatic attraction. The same process was followed when using L-lysine (>98% Sigma Aldrich), Poly-D-Lysine (Mw. 1000-5000 Da, Sigma-Aldrich), BSA (>98%, Sigma Aldrich) and Poly-L-Lysine with a higher molecular weight (Mw. 1000-5000 Da).

Characterization and TEM – EELS measurements:

The mean particle size and morphology of the NIR-NPs were characterized by transmission electron microscopy (TEM) using a FEI Tecnai T20, operated at 200 kV. Zeta potential values of NPs dispersions were obtained by photon correlation spectroscopy measurements (90 Plus Brookhaven Particle Size Analyzer, Brookhaven Instruments Co, Holtsville, NY, USA).

The rest of TEM studies (EDS, HRSTEM-HAADF imaging and EELS) have been developed using a FEI Titan Low-Base microscope, working at 80 kV. This microscope is equipped with a Cs probe corrector, a monochromator and ultra-bright X-FEG electron source. For the low-loss studies, the energy resolution was below 200 meV (even 150 meV for some of the cases) and the spectra were collected in STEM mode, using spectrum-image mode.[36] The convergence and collection angle were 25 and 15 mrad., respectively. The tail of the zero-loss peak has been removed using a power law subtraction method.[37]

EELS Simulations:

EELS simulations have been performed in the discrete dipole approximation (DDA) method as implemented in the DDEELS code[26,38]. The optical constant of bulk metals have been taken from tabulated data.[39] A dielectric constant of 4.20 has been taken for $Si_3N_4$.

**Acknowledgment**

The TEM works have been conducted in the Laborario de Microscopias Avanzadas (LMA) at the Instituto de Nanociencia de Aragon (INA) - Universidad de Zaragoza (Spain). Financial support from the EU thanks to the ERC Consolidator Grant program (ERC-2013-CoG-614715, NANOHEDONISM) and the People Program (CIG-Marie Curie Actions, REA grant agreement n° 321642) are gratefully acknowledged, as well as from the Spanish Ministerio de Economia y Competitividad (FIS2013-46159-C3-3-P). Some of the research leading to these results has received funding from the European Union Seventh Framework Program under Grant Agreement 312483 - ESTEEM2 (Integrated Infrastructure Initiative – I3).This research used resources of the "Plateforme Technologique de Calcul Intensif (PTCI)" (http://www.ptci.unamur.be) located at the University of Namur, Belgium, which is supported by the F.R.S.-FNRS under the convention No. 2.4520.11. The PTCI is member of the "Consortium des Équipements de Calcul Intensif (CÉCI)" (http://www.ceci-hpc.be).